 \documentstyle[aps,twocolumn,prb,epsf]{revtex}

\begin{document}

\draft

 \twocolumn[\hsize\textwidth\columnwidth\hsize\csname@twocolumnfalse\endcsname

\title{Landau Theory of the Phase Transitions in Half Doped Manganites:
Interplay of Magnetic, Charge and Structural Orders}
\author{Fan Zhong$^{1,2}$, and Z. D. Wang$^{1}$}
\address{$^{1}$Department of Physics, University of Hong Kong, Hong Kong, People's Republic of China\\
$^{2}$Department of Physics, Zhongshan University, Guangzhou 510275, People's Republic of China}
\date{\today}
\maketitle

\begin{abstract}
The order parameters of the magnetic, charge and structural orders at
half-doped manganites are identified. A corresponding Landau theory of the
phase transitions is formulated. Many structural and thermodynamical
behaviors are accounted for and clarified within the framework.
In particular, the theory provides a unified picture for the scenario of
the phase transitions and their nature with respect to the variation of the
tolerance factor of the manganites. It also accounts for the origin of the
incommensurate nature of the orbital order and its subsequently accompanying
antiferromagnetic order.
\end{abstract}


\pacs{PACS number(s): 71.10.-w, 75.10.-b, 75.30.Kz, 75.80.+q}

\preprint{ZHONG Fan et al}

 ]


The discovery of ``colossal'' magnetoresistance has stimulated a
renaissance of interest in doped rare--earth manganese. \cite{wollan,cmr}
Intensive investigation has revealed a diversity
of novel phenomena due to the complex interplay among magnetic, charge,
orbital and structural orders. A particular relevant issue is the competition
between magnetic and charge orders in half doped manganites. It poses a great
challenge to theorists upon how to deal with the strong correlation in models
with magnetic, orbital and lattice degrees of freedom. Here we formulate a
Landau theory of phase transitions based on the symmetry of the system in an
attempt to understand a variety of sometime controversial structural and
thermodynamical behaviors. Although it may be argued to be only a mean field
theory which is incorrect at critical points, the structural information,
among others it affords is robust. And it is the order parameters that
exhibit singularity at the critical points.

\begin{figure}
\epsfysize 2.3in \epsfbox{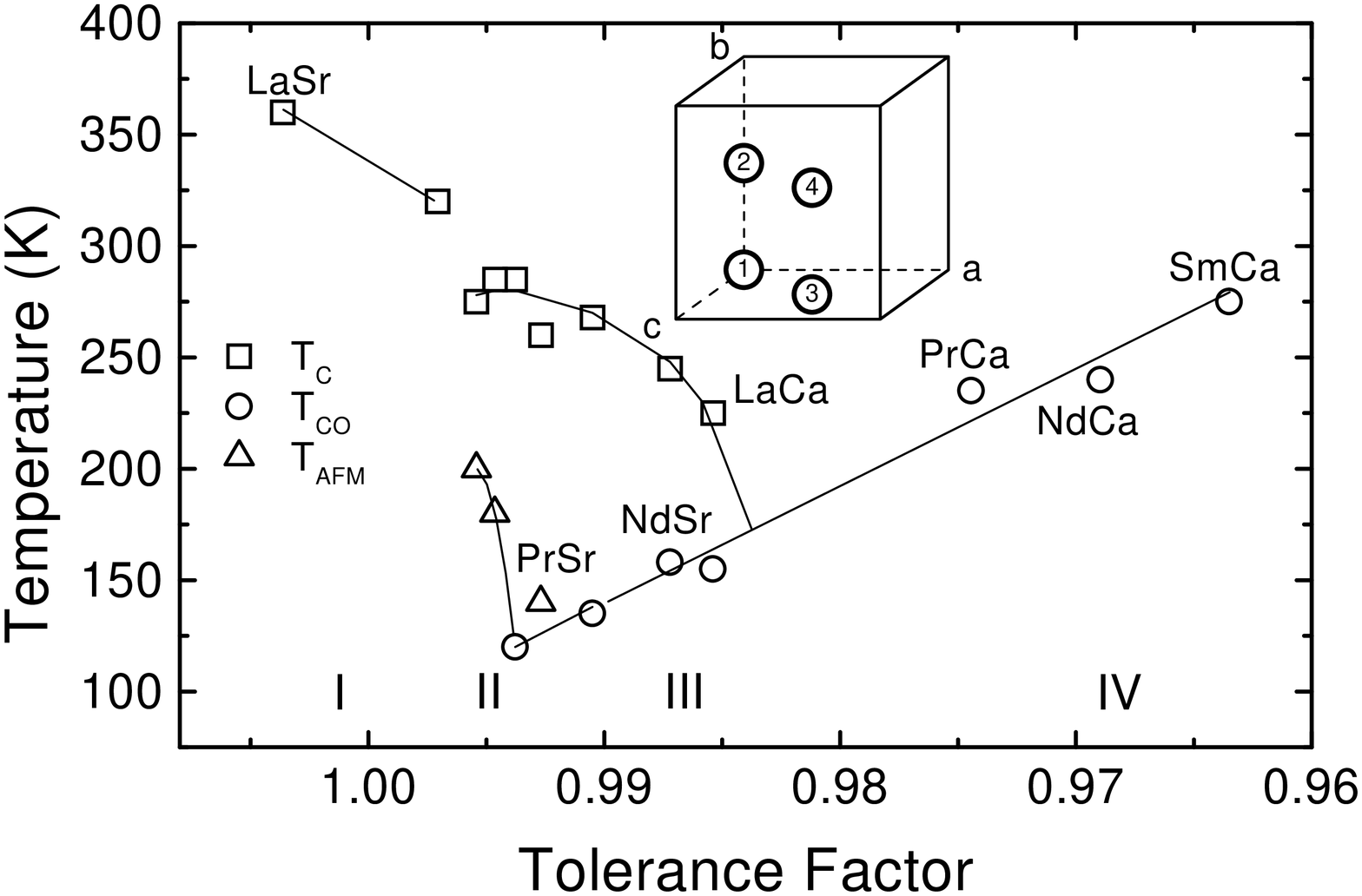}
\caption{Experimental phase diagram of half-doped manganites. Lines are only
a guide for eyes. Sources of the data are given in the text. The inset
displays the elementary unit cell of $Pnma$ and the numbering of the four Mn ions.}
\label{tol}
\end{figure}

The most prominent charge ordered (CO) behavior in perovskite manganites
concerns with those doped at 0.5. These systems exhibit several classes
depending on the tolerance factor of the resultant structure
(see Fig.~\ref{tol}). For La$_{0.5}$Sr$_{0.5}$MnO$_3$ with small distortions,
which we classify as Class I hereafter though no charge order appears (neither
does Class II below), a paramagnetic (PM) to ferromagnetic (FM) transition
occurs at $T_C \sim 360$K. When La is replaced by a smaller ion Nd, $T_C$
decreases with the tolerance factor. At 40 percent of Nd or so, an
intermediately distorted class II shows up in which the FM phase transforms
at a lower temperature $T_{AFM}$ to a metallic A-type antiferromagnetic
(AFM) state \cite{kawano97,moritomo97}. As more La is replaced by Nd, \cite{moritomo97}
a new class III sets in which displays a CO CE-type AFM state
below $T_{CO}$. \cite{wollan} Similar behavior has been reported in La$_{0.5}$Ca$_{0.5}$MnO$_3$.
\cite{radaelli,radaelli97}
Pr$_{0.5}$Sr$_{0.5}$MnO$_3$ is somehow special. It was initially reported to be
CO, \cite{knizek} but later only A-type AFM order was found. \cite{kawano97} However,
Comparing the transport behavior of
Pr$_{0.5}$Sr$_{0.5}$MnO$_3$ with (La$_z$Nd$_{1-z}$)$_{0.5}$Sr$_{0.5}$MnO$_3$
of $0<z\leq0.4$, one finds similar behavior. Only the latter's resistivity
levels off to a presumed metallic state slightly slowly after a jump at $T_{CO}$
or $T_{AFM}$. So the boundary between the CO and A-type AFM states seems to
be not so clear-cut: there may be a transition from the CO state to the
A-type AFM state. For the most distorted class
IV such as Pr$_{0.5}$Ca$_{0.5}$MnO$_3$ and Nd$_{0.5}$Ca$_{0.5}$MnO$_3$, no FM
order appears. \cite{jirak} The PM phase changes directly into a CO state
below $T_{CO}$ and then global AFM ordering shows up at a lower
temperature ($\sim 150$K). The CO tendency in Class IV even extends
to lower doping, although with a pseudo-CE-type structure due to the excess
electrons. \cite{jirak,yoshizawa95} 

The PM to FM transition is continuous. All the other transitions shown
in Fig.~\ref{tol} are of first order with hysteresis. The most striking feature
of the CO state is that it can be melt by an external magnetic field
\cite{kuwahara95}, pressure \cite{moritomo97} or electric field \cite{asamitsu97}
or even by x-ray \cite{kiryukhin} or light irradiations \cite{miyano} into a
FM state, indicating the competition between them. The required magnetic field
to melt the CO state in Class IV is almost twice as large as that in Class III.
\cite{tokunaga,xiao} In this paper, we shall concentrate on the
peculiar phenomena associated with CO. The lattice structure of interest is
orthorhombic with a space group $Pnma$ (see Fig.~\ref{tol} inset), which is well characterized
and the
most common in doped manganites. There are some scatters in the reported
structures. This is understandable because of the small distortion of the
perovskite structure, which is also sensitive to the preparation conditions.
We develop below a Landau theory of the phase transitions in these systems
and show that many structural and thermodynamical behaviors are closely
correlated within the framework. In particular, besides the magnetic,
charge and structural/orbital pattern, primary features are a simple picture
that unifies the three classes along with the nature of the transitions
involved and the origin of the incommensurability of the orbital order,
CE-type AFM order and the melting of CO state.

We start with the magnetic transition. A FM phase transition is associated
with a wave vector at ${\bf k}={\bf 0}$ or the $\Gamma$ point. On the other
hand, the CE-type AFM structure is described by ${\bf k}=(00\frac{1}{2})$
and ${\bf k}=(\frac{1}{2}0\frac{1}{2})$ for Mn ions at positions 1,2 and 3,4,
respectively. In La$_{0.5}$Ca$_{0.5}$MnO$_3$, the FM to AFM
transition is found to be accompanying the incommensurate (IC) to (nearly) commensurate
orbital ordering transition. \cite{radaelli} In Class IV, the CE-type AFM
order can appear separate from the charge and structural orders,
which produce a natural configuration for the CE-type AFM structure (see
below). Thus the AFM order is regarded as originating from the charge and
orbital orders. For simplicity, we leave it towards the end of the paper.

The PM to FM transition is described by an order parameter $M$ representing
the average magnetization over the system. As pointed out in our previous
work, \cite{zhong99} all the irreducible representations (IR's)
at ${\bf k}={\bf 0}$ of $Pnma$ are one dimensional. So we choose $M$ as a
scalar representing the component that carries the IR $\tau^5$ responsible
for the transition. \cite{kovalev} Although it may couple to other magnetic
configurations such as A-type or C-type AFM orders of the same IR, we can
eliminate such modes if present and write the free energy simply as
\begin{equation}
F_M=\frac{1}{2}a_1M^2+\frac{1}{4}b_1M^4,
\label{fm}
\end{equation}
with $a_1=a_{10}(T-T_1)$, and $a_{10}$ and $b_1$ depending only weakly on
the temperature $T$. Eq.~(\ref{fm}) describes a continuous phase transition
at $T=T_1$ with $M=\sqrt{-a_1/b_1}$.

Next we move on to the CO phase transition. A prominent feature that
indicates the existence of the CO state is the appearance of the
superlattice diffraction spots characterized by a wave vector
${\bf k}=(\frac{1}{2}00)$ or the $X$ point in the $Pnma$ setting, which is
adhered to throughout this paper. However, such spots associate more with
lattice modulations than with the charge order that is characterized by
${\bf k}=(100)$ or $\Gamma$. Direct charge ${\it and}$ orbital orders are
first detected by x-ray resonant scattering techniques in doped
La$_2$SrMnO$_4$. \cite{murakami1} So to describe the transition, we need two
order parameters.

On the one hand, the charge order is characterized by an order parameter
$C=\xi_1+\xi_2-\xi_3-\xi_4$ with $\xi_i$ the occupancy possibility of site
$i$ (Fig.~\ref{tol} inset). $C$ is maximized if $\xi_1=\xi_2=-\xi_3=-\xi_4$,
and so a nonzero $C$ produces the observed CO pattern. By considering the
permutations of the four sites under the symmetry operations of the $Pnma$
group, it can be shown that $C$ transforms also as the IR $\tau^5$ of $Pnma$
at $\Gamma$. Accordingly, the free energy of the CO transition is given by
\begin{equation}
F_C=\frac{1}{2}a_2C^2+\frac{1}{4}b_2C^4,
\label{fc}
\end{equation}
where $a_2=a_{20}(T-T_2)$, and $a_{20}$ and $b_2$ are constants. Below $T_2$
charge order appears and the symmetry of the structure is lowered to $P2_1/m$. \cite{stokes}

On the other hand, the structural modulation at ${\bf k}=(\frac{1}{2}00)$ is
described by one of the two IR's of the wave vector, namely, $X_1$ and $X_2$,
which are both two dimensional. \cite{kovalev} Accordingly, the structural
transition is characterized by a two-dimensional order parameter ($\eta_1,\eta_2$).
The physical meaning of the order parameter may be the displacement of the
Mn$^{4+}$O$_6$ octahedra as modeled by Radaelli and coworkers to account for
the diffraction patterns. \cite{radaelli97} It can be shown that
$\eta_2$ may represent an arbitrary linear combination of the $x$ and $z$
components and $\eta_1$ the $y$ component of such displacements. Such a
displacement pattern is consistent respectively with an orbital configuration
of $d_{3x^2-r^2}$ and $d_{x^2-y^2}$, which, when propagate half a period to
$X$, switch to $d_{3z^2-r^2}$ and $d_{y^2-z^2}$, respectively. It may be
possible to choose alternatively an orbital basis such as $d_{3x^2-r^2}$ and
$d_{y^2-z^2}$ and then represent the order parameter as the long-range order
of a certain orbital which assumes a certain angle in the orbital space. We
just note in passing that a certain displacement pattern corresponds to
some orbital order.

Notice that the frequently observed $P2_1/m$ symmetry can only arise from
the IR $X_1$ \cite{stokes}. Therefore, the
free energy for this structural transition is
\begin{eqnarray}
F_{\eta} = && \frac{1}{2}a_3(\eta_1^2+\eta_2^2)+\frac{1}{4}b_3(\eta_1^4+\eta_2^4)+\frac{1}{4}d(\eta_1^2+\eta_2^2)^2 +\nonumber\\
&& \kappa \left(\eta_1 \frac{\partial{\eta_2}}{\partial{x}}-\eta_2 \frac{\partial{\eta_1}}{\partial{x}}\right)+\frac{\sigma}{2} \left[(\nabla \eta_1)^2+(\nabla \eta_2)^2\right],
\label{fe}
\end{eqnarray}
where again $a_3=a_{30}(T-T_3)$, and $a_{30}$, $b_3$, $d$, $\kappa$ and
$\sigma$ are constants.

A peculiar feature of Eq.~(\ref{fe}) is the appearance of the Lifshitz
invariant (the $\kappa$ term), which frequently leads to IC
modulations. \cite{lifshitz,toledano} Many characteristic features of an IC
transition has been observed in the CO and structural transition in manganites.
\cite{radaelli97,mori99} Therefore, the IC nature of the structural/orbital
(but not charge) order has its origin in the Lifshitz invariant of the $X$ point.
Nevertheless, we shall neglect this IC feature of the structural modulation
below for simplicity and focus on its interplay with the charge and magnetic
orders. This is partly justified by the fact that commensurate structure is
also frequently observed in the same experiments that display the reverse.
\cite{radaelli97} In this case, Eq.~(\ref{fe}) then exhibits two possible
phases below $T_3$. One has only one nonzero component equal to
$\pm \sqrt{-a_3/(b_3+d)}$ and so its symmetry is $P2_1/m$ \cite{stokes}
if $b_3+d>0$ and $b_3<0$. The other satisfies
$\eta_1=\pm \eta_2=\pm \sqrt{-a_3/(b_3+2d)}$ and belongs to $Pm$ symmetry
\cite{stokes} when $b_3+2d>0$ and $b_3>0$.

\begin{figure}
\epsfysize 2.3in \epsfbox{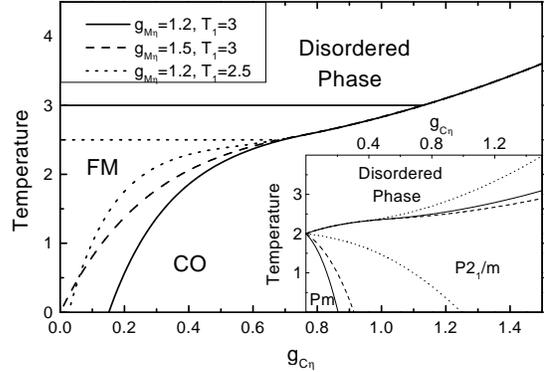}
\caption{Theoretical phase diagram of the coupled magnetic, charge and
structural transitions. The parameters are
$a_{10}=a_{20}=a_{30}=b_1=b_2=b_3=1$, $T_1=3$, $T_2=2.5$, $T_3=2$ and
$g_{M\eta}=1.2$ (solid lines). The dashed line corresponds to
$g_{M\eta}=1.5$, while the dotted lines to $T_1=2.5$. Inset: phase diagram
of the charge-structural transition. The dashed lines represent $b_3=2$
while the dotted ones $d=-0.3$.}
\label{pdmce}
\end{figure}

Coupling of the charge to the structural degrees of freedom can be readily
found by noting that $\eta_1^2-\eta_2^2$ transforms as the same IR as $C$,
and so the simplest coupling between them is
\begin{equation}
F_{C\eta}=g_{C\eta} C(\eta_1^2-\eta_2^2),
\end{equation}
where $g_{C\eta}$ is a measure of the coupling. Assume $g_{C\eta}$ and $C$
are positive without loss of generality. It is transparent then that the
coupling will favor the ordering of $\eta_2$ once the charge is ordered,
since the transition point for $\eta_2$ is now elevated to
$a_3-2 g_{C\eta} C$, while that for $\eta_1$ lowered to $a_3+2 g_{C\eta} C$.
This explains the reason for the often observed $P2_1/m$ symmetry.
\cite{radaelli97} Possibility for ordering of both $\eta_1$ and $\eta_2$
still exists, which may account for the absence of the $2_1$ screw axis in
Sm$_{0.5}$Ca$_{0.5}$MnO$_3$, which should then be of $Pm$ symmetry, though
it was preferred to be $P2mm$ or $Pmmm$. \cite{barnabe} This scenario is
confirmed from the phase diagram illustrated in the inset of
Fig.~\ref{pdmce}. It is seen that as $b_3$ increases, the boundary of the
$Pm$ phase moves to the right, reducing the region of the $P2_1/m$ phase.
If $b_3<0$, only the $Pm$ phase exists, while for $d<0$, the $Pm$ phase
extends far to the right. Note that a large $b_3$ or small $d$ means a
large ``lock-in'' term [the $b_3$ term in Eq.~(\ref{fe})] that tends to
suppress the incommensurability. \cite{toledano} This seems to be consistent
with the observation that more distorted systems such as Sm and Gd tend to
be more stoichiometric and so commensurate. \cite{barnabe} We remark that
the transition from $P2_1/m$ to $Pm$ phase may accompany or be hidden by the
IC to commensurate transition and PM to CE-type AFM transition. 

Since $M$ changes sign by time reversal, the only possible couplings of the
magnetic to the charge and structural transition are bi-quadratic, i.e.,
\begin{equation}
F_{MC\eta}=\frac{1}{2}g_{M\eta} M^2 (\eta_1^2+\eta_2^2)+\frac{1}{2}g_{MC}C^2 M^2,
\label{fmce}
\end{equation}
where both coupling coefficients are positive due to the competing orders.

Equations (\ref{fm}-\ref{fmce}) constitute our theory of the magnetic and
charge and structural or orbital transitions. Instead of going into detailed
estimations of the various coefficients in the model, we are content here
with global features that are believed to be relevant to
the parameter regime of real materials. To this end, we study of a simplified
version of the theory, in which we have taken $\eta_1=0$, i.e., disregarded
the possibility of the $Pm$ symmetry,
relabeled $b_3+d$ as $b_3$, and neglected the bi-quadratic coupling between
the charge and magnetism, since their coupling to the structural order
results in a lower-order $CM^2$-type coupling. This implies that long-range
charge order always accompanies with structural order. 

\begin{figure}
\epsfysize 2.3in \epsfbox{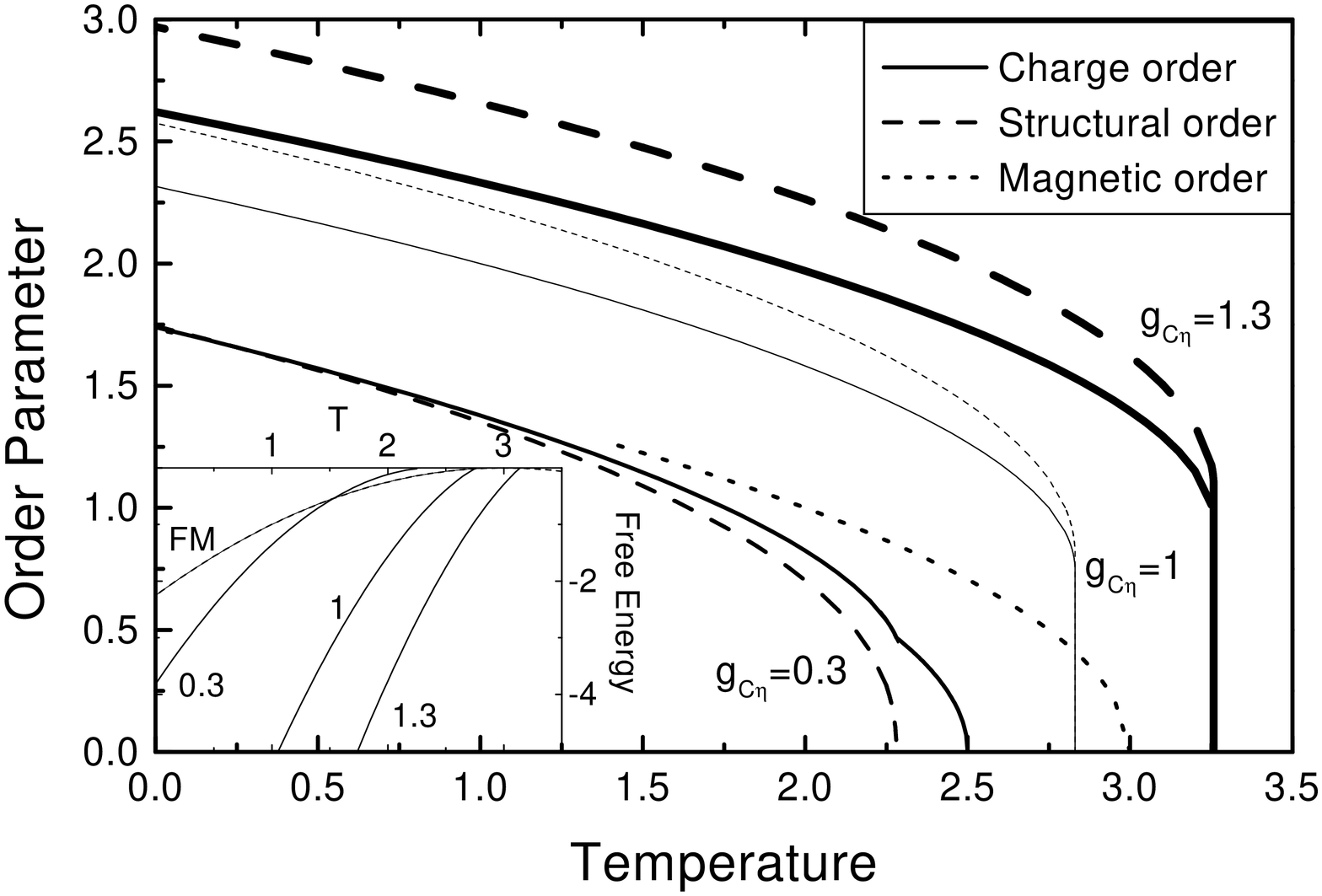}
\caption{Order parameter vs temperature $T$. Thick lines correspond to
$g_{C\eta}=1.3$ while thin and mediate lines $g_{C\eta}=1$ and
$g_{C\eta}=0.3$, respectively. The inset shows the corresponding free energy
from bottom. The upper curve is for the pure FM phase [Eq.(\protect\ref{fm})]. Other parameters are the same as the solid lines in Fig.\protect\ref{pdmce}. Overlaps in $g_{C\eta}=0.3$ imply metastable region.}
\label{mceop}
\end{figure}
Figure \ref{pdmce} displays a generic phase diagram for several
sets of the parameters in arbitrary units. It shows the
strong coupling case $g_{M\eta}^2>b_1b_3$ between magnetism and lattice so
that no mix magnetic and charge orders appears. It can be seen that as
$g_{C\eta}$ increases, the sequence of phase transitions changes from a
PM $\rightarrow$ FM to a PM $\rightarrow$ FM $\rightarrow$ CO then to a
PM $\rightarrow$ CO phase directly as $T$ is lowered. The real situations as
in Fig.~\ref{tol} are of course not simply only a variation of $g_{C\eta}$,
but Fig.~\ref{pdmce} exhibits in a simple way the relevance of the theory.
Figure \ref{mceop} shows the variation of the order parameters for three
different $g_{C\eta}$. It is seen that the transition to CO phase is
discontinuous. The variations of $C$ and $\eta$ are similar, but their
coupling is not necessarily linear. \cite{zimmermann} Note that the CO and
the structural transitions are so strongly coupled that they take place at a
single transition temperature. Although the CO can still appear preceding
the structural order when $g_{C\eta} <1$, the reverse is not true, namely,
$\eta=0$ if $C=0$. Therefore the structural or orbital order is driven by
the charge order. The inset plots the
corresponding free energy vs temperature, showing that the larger the
$g_{C\eta}$, the lower the free energy, and so the bigger the magnetic field
that lowers the upper curve [Eq.~(\ref{fm})] by $HM$ to melt the CO state,
being in agreement with experiments. This indicates that the magnetic field
acts in melting the CO state more than the AFM state. An evidence for this
is that above the CE-type AFM ordering temperature, no AFM for the field
to melt. This also justifies the separate treatment of the AFM order.

Finally we discuss briefly the transition to the CE-type AFM order. Noting
that in La$_{0.5}$Ca$_{0.5}$MnO$_3$ with the standard CE-type state, the
magnetic moments lie in the $a$-$c$ plane \cite{radaelli97}, we may choose
the $x$ and $z$ components of the AFM vectors
${\bf L}_1 = \mbox{{\boldmath $\mu$}}_{1}-\mbox{{\boldmath $\mu$}}_{2}$
and ${\bf L}_2 = \mbox{{\boldmath $\mu$}}_{3}-\mbox{{\boldmath $\mu$}}_{4}$
as order parameters, since they transform as the two components of the
two-dimensional IR's $\tau^1$ and $\tau^1$ combining with its complex
conjugate $\tau^3$ associated respectively with the wave vectors
$(00\frac{1}{2})$ and $(\frac{1}{2}0\frac{1}{2})$, \cite{kovalev}
where {\boldmath $\mu_i$} is the magnetic moment of ion $i$. The two IR's
rather than one makes the transition discontinuous. \cite{toledano}
Both IR's give a magnetic symmetry of $P_b2_1/m$ and a structural one of
$P2_1/m$ which is identical with the CO and structural transitions. \cite{stokes}
The compatibility of the orbital and magnetic patterns implies an enhancement
of both transitions and can be described by a coupling
$(\eta_1^2+\eta_2^2)(L_{1\alpha}L_{1\beta}+L_{2\alpha}L_{2\beta})$
with a {\it negative} coupling constant, where $\alpha$ and $\beta$
denote $x$ or $z$. As a result, ordering of one kind of the orders enhances
the other, leading, for instance, to the accompanying of the AFM order with
the IC to commensurate orbital order \cite{radaelli,mori99} as the AFM
order promotes the contribution of the lock-in term, and to the increasing
of resonant x-ray scattering intensity of charge and orbital orders as AFM
ordering. \cite{zimmermann}

In conclusion, we have developed a Landau theory for the coupled phase
transitions in half-doped manganites through the identification of the
order parameters for the FM, CE-type AFM, CO and structural or orbital
orders. The theory provides a unified picture for the scenario of the
phase transitions and their nature with respect to the variation of the
tolerance factor of the manganites via the symmetry-adapted coupling
among the degrees of freedom. Many peculiar phenomena of half-doped
manganites result from the interplay between the FM or A-type AFM and
CO states, the CE-type AFM sets in only as a secondary factor. So an
applied magnetic field primarily melts the CO state. The theory also
accounts for the origin of the IC nature of the orbital order and its
subsequently accompanying AFM order. As a phenomenological theory, it can
make direct contact with the experimental results especially the symmetry
of the involved structures that sensitively influence the transport behavior
in manganites. Experimental clarifications are desirable of the symmetry of
the CO state and its relation to the oxygen stoichiometry and
commensurablity, possible structural transition from $P2_1/m$ to $Pm$, and
the relation between A-type AFM and CO states.

This work was supported by a URC fund and a CRCG grant at the University of Hong Kong.

\end{document}